\documentclass[fleqn,12pt]{wlscirep}
\usepackage[utf8]{inputenc}
\usepackage[T1]{fontenc}
\usepackage{lscape}
\usepackage{pdflscape}
\usepackage{mathtools}
\usepackage{cuted}
\usepackage{mathtools,bbm}
\usepackage{makecell} 

\usepackage[caption=false,font=footnotesize]{subfig}
\graphicspath{{./images/}{./images/stubborn/}{./images/2type/},{./images/final_res/}}

\title{Navigating Gender Disparities in Communication Research Leadership: Academic Recognition, Career Development, and Compensation}
\author[1*,+]{Diego F. M. Oliveira}
\author[2*]{Qian Huang}
\affil[1]{School of Electrical Engineering and Computer Science, University of North Dakota, Grand Forks, ND, USA.}
\affil[2]{Edward R. Murrow College of Communication, Washington State University, Pullman, WA, USA.}

\affil[+]{diegofregolente@gmail.com}

\affil[*]{These authors contributed equally to this work}

\begin{abstract}
This study investigates gender disparities in communication research, analyzing citation metrics, authorship patterns, team composition, and faculty salaries. Using data from 62,359 papers across 121 communication journals, we observe growing female representation among authors, yet citation gaps remain. Sole-authored papers by women receive fewer citations than men’s, especially in smaller teams. Team composition analysis highlights gender homophily, with single-gender teams being more prevalent. In leading U.S. communication journals, female authors are underrepresented and face citation disadvantages favoring male authors. Salary data from prominent U.S. public universities reveal lower earnings for female Assistant Professors, though gaps diminish at senior ranks. These findings underscore the need for fostering gender equity through inclusive collaborations, fair citation practices, and equitable compensation.
\end{abstract}
\begin{document}

\flushbottom
\maketitle

\thispagestyle{empty}

\section*{Introduction}

In recent years, there has been a noticeable surge in interest and attention focusing on exploring the intersectionality of diversity, equity, and inclusion (DEI) within different areas \cite{west2013role,mairesse2015does,holman2018gender,knobloch-westerwick2013matilda,wang2024systematic,xu2024longitudinal}. This growing interest reflects a broader societal recognition of the importance of understanding and addressing issues related to diversity, equity, and inclusion across fields.

Gender disparities in academia take many forms, including but not limited to matrices measuring academic performance. Gendered compensation is another essential indicator of structural disparity.
Existing research on gender equity and diversity in the field of communication research has focused on authorship distributions, citations, women's representation in conferences, division leadership, and the temporal progress of gender equity and parity over time \cite{wang2021gendered,braun2023gender,trepte2020national}. Following this route, in this paper, we expand the indices of gender equity and representation to other essential components yet to be examined for the field of communication research, such as team composition and salary, to offer a comprehensive overview of gender equity and diversity among communication scholars.

Although communication science has been seen as fragmented with multiple hyper-specialized topics and niche areas, research shows closer connections and more frequent collaborations across subfields of communication\cite{song2020less,tenenboim2020speaking}. Therefore, we include research in all subfields of communication science rather than focusing on certain ones. This approach allows us to obtain a more holistic view of gender inequity in the field of communication science, providing evidence and insight into improving gender equity in the entire field. Therefore, we included journals across all disciplines of communication research, including communication, journalism, and advertising.

In this paper, we present a detailed investigation of Diversity, Equity, and Inclusion (DEI) within teams and leadership roles in the field of communication. Leadership roles, such as sole authorship and first authorship in scientific publications, despite the small number, significantly inform research cultures, recruitment strategies, promotion practices, and mentoring dynamics. To address this issue, we introduce a new set of data using an inaugural comprehensive analysis of women in various leadership roles within the field of research communication.  Despite some progress, gender disparities persist in critical domains, particularly in participation in research collaborations and the recognition of scientific contributions \cite{xiao2018characterizing,williford2023diversity,xiao2023gender,kwiek2021gender,sa2020gender,braun2023gender}. This underscores the pressing need to address these inequities and ensure equal opportunities for women in the field of communication research \cite{schatto2023two}.

The underrepresentation of women in academic research is a multifaceted issue with roots in various societal, institutional, and cultural factors. Although the exact reasons for this phenomenon are not fully understood, researchers have identified several contributing factors. A significant factor is the persistence of traditional gender roles\cite{braun2023gender}, which often place greater domestic and caregiving responsibilities on women, restricting their ability to fully invest in academic pursuits \cite{alesina2013origins}. Sexism within academic and professional settings can create barriers for women, affecting their career advancement and access to leadership positions \cite{wu2017gender}.

Inadequate mentorship opportunities also act as a major hurdle in women's progress and development in their academic career. Studies have shown that women, especially in STEM fields, have experienced difficulties in identifying mentors who can provide guidance and support throughout their academic and professional journey \cite{berg1983men, ibarra2010men,xiao2018association}. This lack of mentorship for women probably results in struggles to navigate the complexities of academia and challenges in advancing their careers \cite{braun2023gender}. Research suggests that positive interactions with faculty members of the same sex can significantly impact students’ academic success\cite{carrell2010sex}. In the absence of women in academic leadership roles, aspiring women researchers may find it challenging to envision themselves in similar positions, perpetuating the cycle of underrepresentation of women. 

Moreover, gender disparities in recognition and rewards exacerbate this issue. Women have been absent from prestigious awards and promotions, despite their significant research contributions \cite{ma2019women}. Deficient acknowledgment of women not only undermines their professional achievements but perpetuates a cycle of gender inequality within academic institutions. These challenges, as a result, enlarge the gender disparities in salary and promotion timelines. In particular, women often receive lower salaries than their male counterparts and experience longer time for promotions to higher academic ranks \cite{ceci2014women, takahashi2011gender}. Furthermore, the distribution of faculty across different ranks highlights the stark gender disparities within academia. Women are disproportionately concentrated in lower-ranking positions, with significantly fewer holding full professorships compared to their male counterparts \cite{Chevalier2020report}. This "leaky pipeline" phenomenon underscores the systemic barriers faced by women in advancing their careers within academia \cite{berg1983men,cotter2001glass}. These disparities reflect systemic biases and contribute to the perpetuation of gender inequality within academia. Meanwhile, the practice of author sorting in scientific publications in fields - such as Economics - can greatly disadvantage women, leading to diminished credit for their contributions compared to male coauthors \cite{sarsons2015gender, mcdowell1999cracks, mcdowell1992effect}. Known as the "Matthew Effects," this phenomenon further marginalizes women in academic research, hindering their professional and personal advancement.

Taken together, the underrepresentation of women in academic research is a complex issue informed by a myriad of societal, institutional, and cultural factors. Addressing this disparity requires concerted efforts to dismantle systemic biases, provide equitable opportunities for mentorship and career advancement, and foster a supportive and inclusive academic environment for women. 
To evaluate gender imbalances in the communication field, our study undertook an analysis centered on the representation of men and women within research teams and high-impact publications. We conducted an analysis of the composition of research teams featured in peer-reviewed publications, categorizing them based on the gender distribution of authors. These categories include sole authors, teams comprised solely of male authors, teams comprised solely of female authors, teams with an equal number of male and female authors, teams with more female authors than male authors, and teams with more male authors than female authors. Our findings reveal that women make up 50.40\% of the authorship. Additionally, our investigation into collaboration patterns indicates a tendency for both male and female researchers to collaborate more frequently with individuals of the same gender, deviating from random chance.

We observed that mixed-gender teams with an equal representation of male and female authors, and those predominantly consisting of males, are less prevalent than expected by chance. On the other hand, teams with a higher proportion of female authors occur more frequently than anticipated by random chance. These trends persist across teams of various sizes, as shown below. Collaborations between female researchers, especially those involving senior female leaders, may foster the advancement of early-career female scientists. However, publications resulting from these collaborations tend to receive fewer citations compared to those with predominantly men and mixed-gender collaborations. Moreover, it is noteworthy that although sole female authors and all-female teams have published in journals with higher impact factors compared to their male counterparts, they still tend to receive fewer citations.

In summary, this paper aims to shed light on the complexities and challenges faced by women in academia, particularly within the field of communication. By presenting an extensive analysis of gender representation in research teams and leadership roles, we provide insights into the persistent disparities that continue to hinder gender equity. The findings underscore the importance of addressing these inequities through targeted interventions that promote diversity, equity, and inclusion in academic environments. Ultimately, our study contributes to the ongoing discussion on gender disparities in academia and emphasizes the urgent need for structural changes to create a more inclusive and equitable landscape for future generations of female researchers.

\section{Data and Method }

We compiled a comprehensive dataset comprising 62,359 papers in communication from 121 different journals between 2004 and 2023, as categorized by the authoritative \textit{Web of Science.} A subset of the dataset comprising 7,017 papers was sampled to examine gender differences in the most influential journals within the field. This dataset served as the foundation for our analysis of gender disparities in authorship and collaboration patterns within the field of communication. To determine the gender of the authors of these publications, we employed a well-established and validated name disambiguation method \cite{Sexmachine,Census}. This method utilizes an extensive list of first names from diverse regions, including Europe, the United States, China, India, and Japan. Each name is assigned a typical gender based on expert assessments and field knowledge. For unisex names or names that couldn't be classified, we referred to the database of the US Census Bureau \cite{Census}. This multi-step approach ensured a high degree of accuracy in gender classification.

To further enrich our dataset, we identified the most important public universities in communication from the National Communication Association. We then emailed each institution and submitted a public records request, asking for the names, titles, and salaries of all faculty members within their communication departments. This information provided a detailed overview of the gender composition and hierarchical structure within these academic units.

Our dataset enabled us to explore the network effects during the selection process and gender homophily within collaboration patterns. By analyzing the co-authorship networks, we could identify the presence of gender-based clustering and homophily, which refers to the tendency of individuals to collaborate with others who share similar characteristics, such as gender. We performed statistical analyses to quantify gender disparities in authorship, citations, and collaboration patterns. This included comparing the observed frequencies of male and female authorship to expected frequencies if gender was unrelated to team composition. We also examined the distribution of authorship across different journals and institutions to identify potential biases or trends in publication practices.

The exploration of network effects and gender homophily provided valuable insights into the intricate relationship between diversity and equity within the academic system. By understanding the mechanisms influencing gender disparities, we aimed to shed light on the factors contributing to these disparities and provide evidence-based recommendations for promoting greater gender equity in communication research. The methods employed in this study allowed for a comprehensive analysis of gender disparities in communication research. By combining a large-scale dataset with robust name disambiguation techniques and detailed institutional data, we were able to provide a nuanced understanding of the factors influencing gender equity within the field.

\section{Differences in Women’s and Men’s Scientific Research - Citations, Impact and Collaboration Patterns}

In the complex landscape of assessing leadership influence and impact within communication research, multiple factors converge to shape perceptions and opportunities. A key consideration in this context is the evaluation of research impact, which often serves as a tangible measure of a researcher's contributions to the field. Citations are commonly used as a metric for this evaluation, representing a form of currency within the academic sphere \cite{li2015big, jacob2011impact, symonds2006gender, balkundi2006ties}.

Funding agencies, as central stakeholders in academia, frequently rely on citation metrics to assess the significance and reach of research efforts. The number of citations a researcher's publications receive is often used as a proxy for the influence and relevance of their work within scholarly discourse. As a result, this metric plays a crucial role in determining the allocation of resources and in shaping academic trajectories. The impact of research, as quantified through citation metrics, significantly influences various aspects of a researcher's academic career. Achievements such as tenure and promotion are often contingent upon the perceived impact of one's scholarly work. Moreover, opportunities to assume leadership roles within the field of communication research may depend on the ability to generate impactful contributions, as evidenced by citation counts.

In the interconnected academic ecosystem, the significance of research impact extends beyond recognition to transdisciplinary influence that can shape perceptions of expertise and authority in the broader scholarly community. Furthermore, the pursuit of impactful research is closely tied to the advancement of knowledge and the dissemination of ideas, highlighting its societal relevance. Understanding the multifaceted dynamics of research impact is essential for navigating the complexities of academic leadership and influence within communication research. By acknowledging the central role of citation metrics in informing perceptions and opportunities, researchers can strategically position themselves to maximize their impact and contribute meaningfully to the advancement of knowledge in the field.

Research in academia is inherently collaborative and presumably performed by teams \cite{wuchty2007increasing}. The synergy of diverse minds and collective effort frequently yields richer insights and more robust findings than individual efforts \cite{yang2022gender}. In this collaborative environment, assessing leadership influence becomes crucial, influencing not only the trajectory of research but also the recognition and advancement of scholars. Proxy measures commonly used to gauge leadership impact include sole authorship and citation counts \cite{national2015enhancing, wang2013quantifying}. In particular, sole authorship often represents a significant degree of autonomy and expertise. Citation counts reflect the resonance and influence of the work within the scholarly community. However, these metrics only partially represent the intricate dynamics at play within research teams.

In various academic disciplines, such as communication, the order of authorship in scholarly papers carries significant implications for leadership and credit allocation. Authorship order entails roles and responsibilities within the research process, indicating contribution of each team member. These authorship positions are most commonly recognized: sole author, first author, and last author \cite{wuchty2007increasing}. First, sole authors are seen as steering all aspects of the research, from conceptualization to execution and dissemination, representing a singular vision and expertise. Meanwhile, first authors are credited for their substantial contributions to the paper and major responsibility for research workload. Their name appears first in the publication, suggesting their key role in content and direction. The last author, often the senior author, takes on a supervisory role, guiding critical publication-related decisions, providing strategic direction to the research, and acquiring funding for the project. These authorship positions not only denote individual contributions but also inform the authors' career progression and scholarly reputation. In addition, papers are commonly cited using the first (or sole) author's name, amplifying the importance of authorship order in shaping scholarly recognition and impact. Ultimately, an author's positioning within the authorship hierarchy will determine promotion and tenure decisions and subsequent strategic collaboration and mentorship within research teams. 

Therefore, another aim of this study is to explore the impact and recognition within teams based on the structure of the team in each publication. We categorize published papers based on team compositions, ranging from sole authorship to collaborations with diverse gender compositions. Specifically, we categorized published papers by (a) sole authorship, (b) an equal number of female and male authors, (c) teams with more female authors than males, and (d) collaborations with more male authors than females.

Through a detailed analysis of these team compositions and their research impact, we seek to gain deeper insights into leadership dynamics within the field of communication and beyond. This allows us to see how team composition contributes to recognition and impact within the field, revealing potential disparities or existing biases. Understanding these nuances in team compositions and their associations with research impact is vital to triangulating insights into fostering a more equitable and inclusive research culture within the field. By acknowledging and addressing any biases or imbalances in credit allocation and leadership recognition, we can create an environment that values diverse contributions and provides equal opportunities for researchers to thrive and advance in their careers.


\begin{table}[t]
\title*{Average Journal's Impact Factor}
\centering
\resizebox{\textwidth}{!}{%
\begin{tabular}{@{}|c|ccc|ccc|@{}}
\midrule
 Team Size & ~  & Male Authors & ~ & ~ &  Female Authors & P-Value\\ \midrule
Sole Author & - & 1.53 & - & - & 1.54 & 0.0317 \\ \bottomrule \midrule
 ~ & Only Male  & Only Female  & P-Value & More Male Authors & More Female Authors & P-Value\\ \midrule
2 & 1.92 & 1.93 & 0.1907 & - & - & - \\
3 & 2.17 & 2.22 & 0.0067 & 2.19 & 2.19 & 0.2706 \\
4 & 2.15 & 2.28 & 0.0008 & 2.23 & 2.23 & 0.09 \\
5 & 2.24 & 2.28 & 0.1258 & 2.35 & 2.35 & 0.5827 \\ \bottomrule
\end{tabular}%
}
\caption{The table presents the average journal impact factors segmented by the gender composition of author teams. It highlights the comparison between sole-authored papers by male and female authors, as well as collaborative works across varying team sizes. The results suggest slight differences in the average impact factors, with female-only teams showing a marginally higher average in most cases. However, the p-values indicate that these differences are not always statistically significant, suggesting that the impact factor is generally comparable across gender compositions, with a few exceptions.}
\label{table3}
\end{table}

\begin{table}[t]
\caption*{Average Number of Citations Received}
\centering
\resizebox{\textwidth}{!}{%
\begin{tabular}{@{}|c|ccc|ccc|@{}}
\midrule
 Team Size  & Only Male  & Only Female  & P-Value & More Male Authors & More Female Authors & P-Value\\ \midrule
1 & 3.49 & 3.36 & 0.016 & - & - & - \\
2 & 4.62 & 4.2 & 0.6646 & - & - & - \\
3 & 5.32 & 4.12 & 0.0004 & 4.85 & 4.56 & 0.1186 \\
4 & 4.91 & 4.37 & 0.6437 & 5.02 & 4.74 & 0.1708 \\
5 & 5.36 & 4.66 & 0.3596 & 5.44 & 4.91 & 0.3583 \\ \bottomrule
\end{tabular}%
} 
\caption{The table shows the average number of citations received by papers based on the gender composition of the author teams. It reveals that sole-authored papers by male authors receive slightly more citations on average than those by female authors, with a statistically significant difference. For teams of three or more authors, papers authored by male-only teams tend to receive more citations than those by female-only teams, but the differences are generally not statistically significant when considering teams with a mix of male and female authors. This suggests that gender-related citation disparities may be more pronounced in sole or small-group authorship scenarios.}
\label{table4}
\end{table}

\begin{table}[t]
\centering
\resizebox{\textwidth}{!}{%
\begin{tabular}{@{}|cc|c|ccccc@{}}
\toprule
\multicolumn{4}{|c}{Number of Authors} \\
\cline{2-3}
\multicolumn{8}{|c|}{Probability of Forming a Team With} \\
\cline{4-8}
 Team Size & Male  & Female  & Only Male (\%) & Only Female (\%) & Same Number of  Male and Female (\%) & More Male (\%) & More Female (\%) \\ \midrule \midrule
2 & 18884 & 18300 & 11.45 & 12.19 & -11.81 & nan & nan \\
3 & 16334 & 16021 & 41.3 & 41.26 & nan & -14.82 & -12.69 \\
4 & 9746 & 10210 & 78.63 & 76.54 & -21.21 & -0.48 & -6.36 \\
5 & 5151 & 5889 & 177.13 & 124.36 & nan & -5.35 & -13.54 \\ \bottomrule
\end{tabular}%
}
\caption{This table presents the percentage of collaborations observed among teams from communication journals in the Web of Science (WoS hereafter) data, compared to the expected collaborations for the overall population. The randomization process was repeated 1000 times. Our normalized measure is based on the observed collaboration frequency between a gender pairing $(f_o)$ and the frequency generated by random teams $(f_r)$, which represents the expected collaboration if gender were unrelated to the decision. If $f_o = f_r$, it indicates that collaborations among an author pairing are independent of gender. When $f_o > f_r$, there is a statistical preference for the team structure, and vice versa when $f_o < f_r$. }
\label{my-table02}
\end{table}

\begin{table}[t]
\centering
\resizebox{\textwidth}{!}{%
\begin{tabular}{@{}|cc|c|ccccc@{}}
\toprule
\multicolumn{4}{|c}{Number of Authors} \\
\cline{2-3}
\multicolumn{8}{|c|}{Probability of Forming a Team With} \\
\cline{4-8}
 Team Size & Male  & Female  & Only Male & Only Female & Same Number of  Male and Female & More Male & More Female \\ \midrule \midrule
2 & 2724 & 1960 & 3.75 & 7.24 & -5.21 & - & - \\
3 & 2804 & 2065 & 22.96 & 37.36 & - & -14.02 & -4.23 \\
4 & 1405 & 1139 & 32.16 & 99.61 & -17.78 & 2.61 & -6.3 \\
5 & 760 & 615 & 139.05 & 148.86 & - & -16.11 & -2.6 \\ \bottomrule \bottomrule \bottomrule \bottomrule
\end{tabular}%
}
\resizebox{\textwidth}{!}{%
\begin{tabular}{@{}|cc|c|cccccc@{}}
\multicolumn{9}{|c|}{Average Number of Citations Received} \\
\cline{4-9}
 Team Size & Male  & Female  & Only Male  & Only Female  & P-Value & More Male Authors & More Female Authors & P-Value\\ \midrule \midrule \midrule
1 & 814 & 554 &  5.56 & 6.2 & 0.9594 & - & - & - \\
2 & 1176 & 658 & 7.71 & 6.56 & 0.9153 & - & - & - \\
3 & 2088 & 1521 & 6.82 & 6.34 & 0.9819 & 7.29 & 7.28 & 0.5325 \\
4 & 759 & 541 & 6.85 & 7.68 & 0.7625 & 7.44 & 7.72 & 0.9228 \\
5 & 543 & 442 & 6.42 & 12.89 & 0.8956 & 7.22 & 7.33 & 0.6501\\ \bottomrule
\end{tabular}%
}
\caption{  The top table displays the percentage of collaborations observed among teams for the most influential journals in communication, as recorded in the WoS data between 2004 and 2023. The values are presented as the ratio of observed collaborations to expected collaborations for the overall population. The bottom table shows the average number of citations received by single-authored papers and teams for the most influential journals in communication. The citation data is based on the number of citations received within the first two years after publication, as obtained from the WoS data.}
\label{table06}
\end{table}

\begin{table}[t]
\centering
\resizebox{\textwidth}{!}{%
\begin{tabular}{|c|c|c|c|c|c|c|c|}
    \hline
    \multicolumn{4}{|c|}{} & \multicolumn{4}{c|}{\textbf{Difference In Citations Between}} \\
    \cline{5-8}
    {\bf Journal Name} & {\bf \makecell{Number of \\Publications}} & {\bf IF} & {\bf \makecell{Percentage of\\ Female}} & {\bf \makecell{Female and\\Male - Sole (\%)}} & {\bf \makecell{Female First and\\Male First (\%)}} & {\bf \makecell{Team with only Female\\and Team with only Male (\%)}} & {\bf \makecell{Teams with More Female\\Than Male (\%)}} \\ \hline
       \textbf{PUBLIC OPINION QUARTERLY} & 754 & 3.4 & 33.1 & 14.64 & -21.94 & -36.81 & -28.09 \\ \hline
        \textbf{POLITICAL COMMUNICATION} & 498 & 7.5 & 36.0 & 11.39 & -25.32 & -27.8 & -36.0 \\ \hline
        \textbf{COMMUNICATION METHODS AND MEASURES} & 133 & 11.4 & 39.0 & 107.59 & -76.97 & -82.07 & -78.4 \\ \hline
        \textbf{INTERNATIONAL JOURNAL OF PRESS-POLITICS} & 431 & 4.8 & 42.6 & -9.54 & 9.05 & -20.27 & -30.71 \\ \hline
        \textbf{JOURNAL OF ADVERTISING} & 619 & 5.7 & 42.8 & -34.85 & 0.87 & 49.35 & 26.4 \\ \hline
        \textbf{DIGITAL JOURNALISM} & 658 & 5.4 & 43.9 & 5.45 & -19.86 & -30.24 & -15.39 \\ \hline
        \textbf{JOURNAL OF COMMUNICATION} & 778 & 7.9 & 44.2 & 14.46 & -2.73 & 6.94 & -2.66 \\ \hline
        \textbf{INTERNATIONAL JOURNAL OF ADVERTISING} & 690 & 6.7 & 45.4 & -5.63 & -14.62 & -6.55 & 26.3 \\ \hline
        \textbf{NEW MEDIA \& SOCIETY} & 1760 & 5 & 46.6 & 14.56 & -3.31 & -3.57 & -4.2 \\ \hline
        \textbf{COMMUNICATION RESEARCH} & 696 & 6.2 & 47.0 & 8.78 & 13.21 & -5.4 & -4.19 \\ \hline
\end{tabular}%
}
\caption{The table presents data from top journals in the field of communication, listing the number of publications, impact factor (IF), percentage of female authors, and the differences in citation percentages between male and female authors. The differences are shown separately for sole-authored papers, first author, teams with only males and only females, and for papers with teams with more females than males. Positive values indicate higher citations for male authors or teams with more male authors. On the other hand, negative values indicate greater number of citations for female authors or teams with more female authors. The journals are listed in descending order of female representation, and the data highlights the gender disparities in citation practices within the field.}
\label{fig_top}
\end{table}

We analyzed a dataset comprising of 62,359 papers in communication from 121 different journals between 2004 and 2023, as categorized by the authoritative Web of Science (WoS) \cite{mukherjee2017nearly}. To determine the gender of the authors of these publications, we employed a well-established and validated name disambiguation method \cite{Sexmachine,Census}. This method uses an extensive list of first names from diverse regions, including Europe, the United States, China, India, and Japan. Each name is assigned a typical gender based on expert assessments and field knowledge. Unisex names are identified, and for names that could not be classified, we referred to the database of the US Census Bureau \cite{Census}.

Our analysis of gender disparities in communication research uncovered two significant findings. First, women constitute 50.40\% of all authors in papers indexed by WoS, reflecting the substantial progress women have made in taking on greater roles in scientific research since the 1970s \cite{parker2021s}. Additionally, our analysis reveals that women's representation in papers is nearly on par with men's, comprising 48.85\% and 48.34\% of authors in journals with the top 10\% and bottom 10\% impact factors, respectively. Moreover, among the 62,359 articles surveyed, women are the majority in 25,825 of them, accounting for 41.41\%.

Second, within the extensive WoS sample, 37.75\% of them are sole-authored, with nearly half (47.71\%) authored by women. Notably, women tend to publish their sole-authored work in journals with slightly higher impact factors compared to men, suggesting that differences in publication rates are not solely attributable to variations in research quality between genders. However, despite this, women tend to receive fewer citations on average, as shown in Tables \ref{table3} and \ref{table4}.

Collaborative efforts are essential to advance team science, and research highlights the significant advantages of diversity within these collaborations. Whether the diversity is demographic, such as ethnicity or gender, or cognitive, involving varied expertise and perspectives, such collaborations tend to produce work that attracts more citations and recognition. For example, Freeman and Huang \cite{freeman2015collaborating} find a positive correlation between the ethnic diversity of authors and the citation impact of research papers. This finding underscores the broader merits of diversity for academic success, including greater opportunities for promotion, increased prestige, and more chances of securing funding. The increased citation rates associated with more diverse collaborations suggest that such teams are more likely to produce influential research \cite{yang2022gender}. This creates a constructive feedback loop: diverse collaborations contribute to a more fruitful and impactful research output and boost the visibility and influence of the work produced. As a result, scholars involved in diverse teams may experience greater academic recognition and career advancement, which in turn encourages further diverse and collaborative research efforts. This dynamic emphasizes the value of fostering diversity in research teams to maximize scholarly impact and improve scientific progress.

\begin{table}[t]
    \centering
    \scriptsize 
    \begin{tabular}{|p{0.28\linewidth}|p{0.2\linewidth}|p{0.2\linewidth}|p{0.2\linewidth}|}
    \hline
        \textbf{University} & \textbf{Number of Assistant Professors / Number of female / \% of female} & \textbf{Number of Associate Professors / Number of female / \% of female} & \textbf{Number of Professors / Number of female / \% of female} \\ \hline
        \textbf{Ohio State Univ.} & 6 / 5 / 83.33\% & 17 / 8 / 47.06\% & 7 / 3 / 42.86\% \\ \hline
        \textbf{Univ. of California - Davis} & 1 / 0 / 0.00\% & 5 / 2 / 40.00\% & 7 / 2 / 28.57\% \\ \hline
        \textbf{Univ. of Michigan} & 5 / 2 / 40.00\% & 7 / 3 / 42.86\% & 10 / 5 / 50.00\% \\ \hline
        \textbf{Univ. of Wisconsin-Madison} & 4 / 2 / 50.00\% & 3 / 1 / 33.33\% & 16 / 8 / 50.00\% \\ \hline
        \textbf{Univ. of Texas at Austin} & 24 / 16 / 66.67\% & 27 / 19 / 70.37\% & 49 / 23 / 46.94\% \\ \hline
        \textbf{Univ. of Washington} & 4 / 2 / 50.00\% & 4 / 2 / 50.00\% & 8 / 7 / 87.50\% \\ \hline
        \textbf{Univ. of Illinois Urbana-Champaign} & 4 / 2 / 50.00\% & 6 / 4 / 66.67\% & 11 / 4 / 36.36\% \\ \hline
        \textbf{Univ. of North Carolina at Chapel Hill} & 3 / 2 / 66.67\% & 9 / 4 / 44.44\% & 7 / 4 / 57.14\% \\ \hline
        \textbf{Purdue Univ.} & 9 / 5 / 55.56\% & 2 / 1 / 50.00\% & 11 / 5 / 45.45\% \\ \hline
        \textbf{Univ. of Minnesota} & 3 / 2 / 66.67\% & 2 / 1 / 50.00\% & 5 / 4 / 80.00\% \\ \hline
        \textbf{Michigan State Univ.} & 4 / 2 / 50.00\% & 8 / 2 / 25.00\% & 9 / 3 / 33.33\% \\ \hline
        \textbf{Univ. of Arizona} & 3 / 3 / 100.00\% & 1 / 0 / 0.00\% & 6 / 1 / 16.67\% \\ \hline
        \textbf{Overall} & 70 / 43 / 61.42\% & 91 / 47 / 51.64\% & 146 / 69 / 47.26\% \\ \hline        
    \end{tabular}
    \caption{Table summarizing the gender distribution of faculty members at various U.S. universities across different academic ranks. For each institution, the table lists the number of Assistant Professors, Associate Professors, and Professors, along with the number and percentage of female faculty members in each rank. The data highlights the representation of women in academic positions at these institutions and provides an overview of gender disparities at different career stages within the faculty ranks.}
    \label{table_number}
\end{table}

\begin{table}[t]
    \centering
    \scriptsize 
    \begin{tabular}{|p{0.2\linewidth}|p{0.22\linewidth}|p{0.22\linewidth}|p{0.22\linewidth}|}
    \hline
        \textbf{University} & \textbf{Average Salary Assistant Professors (male) / Average Salary Assistant Professors (female) / \% difference} & \textbf{Average Salary Associate Professors (male) / Average Salary Associate Professors (female) / \% difference} & \textbf{Average Salary Professors (male) / Average Salary Professors (female) / \% difference} \\ \hline
        \textbf{Ohio State Univ.} & 94,783.77 / 93,335.10 / -1.53\% & 109,571.40 / 107,470.87 / -1.92\% & 232,651.75 / 171,945.66 / -26.09\% \\ \hline
        \textbf{Univ. of California - Davis} & 80,025.0/-/- & 99,550.00 / 101,437.50 / 1.90\% & 121,305.00 / 154,387.50 / 27.27\% \\ \hline
        \textbf{University of Michigan} & 91,284.67 / 93,680.00 / 2.62\% & 128,060.75 / 124,875.00 / -2.49\% & 161,919.40 / 175,325.00 / 8.28\% \\ \hline
        \textbf{Univ. of Wisconsin-Madison} & 102,112.50 / 110,297.50 / 8.02\% & 114,885.50 / 127,359.00 / 10.86\% & 162,262.50 / 162552.12 / 0.18\% \\ \hline
        \textbf{University of Texas at Austin} & 103,781.25 / 98,603.12 / -4.99\% & 131,337.25 / 121,522.66 / -7.47\% & 161,014.50 / 156636.79 / -2.72\% \\ \hline
        \textbf{Univ. of Washington} & 133,290.00 / 133,536.00 / 0.18\% & 133,298.14 / 165,522.00 / 24.17\% & 159,888.00 / 156,218.57 / -2.30\% \\ \hline
        \textbf{University of Illinois Urbana-Champaign} & 90,415.50 / 90,235.50 / -0.20\% & 110,887.50 / 106,931.25 / -3.57\% & 138,673.25 / 136687.69 / -1.43\% \\ \hline
        \textbf{Univ. of North Carolina at Chapel Hill} & 82,750.00 / 87,309.50 / 5.51\% & 106,186.60 / 104,854.75 / -1.25\% & 138,476.67 / 148,256.25 / 7.06\% \\ \hline
        \textbf{Purdue Univ.} & 106,739.67 / 80,190.84 / -24.87\% & 142,045.48 / 121,094.97 / -14.75\% & 142,111.33 / 133,537.41 / -6.03\% \\ \hline
        \textbf{Univ. of Minnesota} & 87,521.00 / 85,270.50 / -2.57\% & 99,196.00 / 99,242.00 / 0.05\% & 128,927.00 / 132,057.80 / 2.43\% \\ \hline
        \textbf{Michigan State Univ.} & 88,000.00 / 94,444.45 / 7.32\% & 98,055.52 / 112,565.55 / 14.80\% & 155,699.70 / 191,906.13 / 23.25\% \\ \hline
        \textbf{Univ. of Arizona} & -/76,515.0/- & 88,842.0/-/- & 134,944.60 / 106,986.00 / -20.72\% \\ \hline
        \textbf{Overall} & 100,139.94/94,520.27/-5.61\% & 113,982.39/116,961.97/2.61\% & 155,351.82 / 155,307.66 / -0.03\% \\ \hline        
    \end{tabular}
    \caption{Table showing the average salaries for male and female faculty members at various U.S. universities, broken down by academic rank (Assistant Professors, Associate Professors, and Professors). For each institution, the table includes the average salary for male and female faculty members at each rank, along with the percentage difference in salaries between the genders. The data provides a comparison of salary disparities across universities and ranks, offering insights into gender-based salary differences within academic positions. Note that some entries are missing because a comparison could not be made for those cases, as the university had only male or female faculty members at that rank.}
    \label{table_salary}
\end{table}

The composition of collaborations and the connections they foster are pivotal in shaping professional networks and mentoring, which in turn play significant roles in reputation building and accessing major opportunities \cite{granovetter2005impact,rivera2010dynamics}. These networks can inform a range of professional outcomes, including career advancement, salary levels, tenure rates, patenting success, grant application, and overall recognition within the scientific community \cite{balkundi2006ties,uzzi2013atypical,sargent2004careers,azoulay2010superstar}. 

In science and academia, social networks are integral to professional success as opposed to incidental. They can offer researchers access to collaborative projects, funding opportunities, and influential contacts that might otherwise be unavailable. A well-established network can enhance a researcher’s visibility, provide access to cutting-edge research, and open doors to prestigious positions and awards. Given the critical role of collaborations, fostering diverse and inclusive networks is essential for promoting innovation and achieving greater impact in communication research. Diverse networks bring together individuals with a wide range of backgrounds, perspectives, and expertise, which can lead to more innovative solutions and a richer exchange of ideas. Such diversity can drive groundbreaking research by integrating multiple viewpoints and approaches, leading to more comprehensive and impactful findings. 
Moreover, inclusive collaborations can significantly enhance the visibility and career prospects of researchers, particularly those from underrepresented groups. Establishing strong mentoring relationships within these diverse networks can provide invaluable support and guidance for early-career researchers. Mentoring can help navigate career challenges, provide professional advice, and build confidence, which is crucial for career development and success.

Investing in the development of robust and diverse professional networks is not just a strategic advantage but a means to foster a more inclusive and dynamic research community. By embracing the power of diverse collaborations and supporting inclusive mentoring practices, the field of communication can advance its research agenda, enrich the academic environment, and contribute to a more equitable and innovative scientific landscape. We undertook a detailed analysis of academic authorship collaborations, examining five distinct authorship types: (a) teams composed exclusively of male authors, (b) teams composed exclusively of female authors, (c) teams with an equal number of male and female authors, (d) teams with a higher proportion of female authors compared to male authors, and (e) teams with a higher proportion of male authors compared to female authors. 

Our study focused on research teams with up to five members, which represent 96.40\% of the total publications.
This approach allowed us to capture a comprehensive view of the gender composition in academic collaborations. By analyzing these diverse authorship configurations, we aimed to uncover patterns and trends related to gender dynamics within academic teams and their potential impact on research output and collaboration effectiveness. To thoroughly assess the representations of female and male authors across team sizes, we utilized a randomization approach to control for gender biases in collaboration patterns. Our study began by compiling a dataset of authorship collaborations from communication journals listed in the Web of Science (WoS) database. We then employed a randomization technique to simulate the null hypothesis that gender does not influence authors' decisions to collaborate. Specifically, we randomly reassigned genders to the authors in our dataset and recalculated the collaboration patterns that would be expected under this null hypothesis. This process was repeated 1000 times to create a robust distribution of expected collaboration frequencies ($f_r$) for various authorship configurations, ranging from teams with two to five authors per paper published.

To ensure a rigorous comparison, we then compared these simulated collaboration frequencies with the actual observed collaboration frequencies ($f_o$) from our dataset. This comparison allowed us to identify any significant deviations between the observed and expected collaboration patterns. By analyzing these deviations, we could determine whether gender had a discernible effect on the composition and size of research teams, and whether certain gender disparities in collaboration were statistically significant. This methodology provided us with a quantitative basis for evaluating the influence of gender on research collaboration, revealing insights into potential biases and contributing to a more nuanced understanding of authorship dynamics in the field of communication studies.
The analysis of collaboration patterns in communication journals reveals distinct gender-based preferences in team composition across different team sizes. For teams of 2, the percentage of Only Male (11.45\%) and Only Female (12.19\%) teams is higher than what would be expected by chance. Conversely, teams with an equal number of males and females are observed less frequently, with a percentage of -11.81\%, indicating a tendency towards single-gender teams rather than mixed-gender pairs. In teams of 3, the percentages of Only Male (41.3\%) and Only Female (41.26\%) teams also exceed random expectations. This suggests that single-gender teams are notably more common than mixed-gender teams, which are less frequent than expected, with observed percentages of -14.82\% for more males and -12.69\% for more females. For teams of size 4, the trend continues with even greater deviations from random expectations. The observed percentages of Only Male (78.63\%) and Only Female (76.54\%) teams are significantly higher than expected, reflecting a strong preference for single-gender teams. Mixed-gender teams with an equal number of males and females are less likely, as indicated by the percentage of -21.21\%; while teams with more females (-6.36\%) or more males (-0.48\%) are also less frequent than randomly anticipated. In teams of size 5, the preference for single-gender teams is even more evident. The percentages for Only Male (177.13\%) and Only Female (124.36\%) teams are greater than random expectations. Teams with more males (-5.35\%) or more females (-13.54\%) are observed less frequently than expected by chance.

These results highlight evident gender-based influence on team composition in communication journals. The consistent pattern across various team sizes indicates a strong tendency towards single-gender teams rather than mixed-gender collaborations. This finding underscores the need for further exploration into the factors driving these gender disparities and their implications for research collaboration dynamics.

Thought leadership in academic communication is often gauged by the influence of an author’s research, which is commonly measured through citation counts. To better understand the impact of gender on research visibility and influence, we conducted an analysis focusing on citation metrics associated with different authorship configurations: sole authors, same-gender teams, and mixed-gender teams. Our analysis employs two widely recognized measures of research impact: the journal impact factor (JIF) (according to the Journal Citation Reports (JCR) database from 2022) and citation count \cite{stringer2010statistical, wang2013quantifying, uzzi2013atypical}.

The journal impact factor (JIF) is a metric designed to reflect the average number of citations that articles published in a journal receive over a specific period \cite{garfield2006history}. This measure is often used as an indicator of a journal’s overall quality and influence within the academic community. However, JIF has its limitations. The metric is defined by size and scope of the research community of a field. Therefore, journals in smaller or more niche or specialized fields might receive JIF scores even if the papers published in the field have a substantial impact within the realm. This inevitable discrepancy may restrain generalizability of insights provided by the JIF. JIF may not fully capture the significance of research published in less prominent journals. Additionally, citation counts, another metric used in our analysis, measure the total number of times a research paper is cited by other works. This measure can offer a more direct indication of individual papers' influence and authors' research impact. However, citation counts can be affected by the authors' publication frequency, the size of their research network, and the field’s citation practices. By analyzing these metrics across different authorship compositions, we aim to uncover any patterns or disparities that might reveal how gender influences research impact, visibility, academic influence, and thought leadership.

Table \ref{table3} presents an analysis of average journal impact factors based on the gender composition of author teams of different sizes. The results reveal several notable patterns. For sole-authored papers, female authors achieve a slightly higher average impact factor (1.54; Table \ref{table3}) compared to their male counterparts (1.53). The p-value of 0.03 indicates that this difference is statistically significant, suggesting that sole-authored papers by female authors tend to be published in journals with marginally higher impact factors.

In two-author teams, the average impact factors are nearly identical for male-only and female-only teams (1.92 and 1.93, respectively), with a p-value of 0.19, indicating no significant difference in gender composition. In contrast, in three-author teams, female-only teams have a higher average impact factor (2.22) compared to male-only teams (2.17), with a p-value of 0.01, highlighting advantages for female-only teams. Within this group, teams with more male authors and those with more female authors have an average impact factor of 2.19, with a p-value of 0.27, showing no significant difference.

For four-author teams, female-only teams again outperform male-only teams with an average impact factor of 2.28 compared to 2.15, and a p-value of 0.0008. In contrast, when comparing teams with a greater proportion of male authors to those with more female authors, both groups have identical impact factors (2.23), with a p-value of 0.09, indicating no significant difference. In five-author teams, the difference in impact factors between male-only (2.24) and female-only (2.28) teams is not statistically significant (p-value = 0.13). Similarly, teams with a majority of male or female authors both have an average impact factor of 2.35, with a p-value of 0.58, indicating no significant difference. The analysis suggests that while female-only teams achieve slightly higher average impact factors than male-only teams, these differences are statistically significant only for sole authors, three-author teams, and four-author teams. For larger teams (i.e., those with more than 5 authors), especially those with mixed or balanced gender compositions, the differences in impact factors are negligible and not statistically significant, indicating that gender composition does not substantially influence a publication's impact factor, although certain instances reveal a performance advantage for female-only teams.

Table \ref{table4} provides an analysis based on the gender composition of authors of teams of different sizes. In the analyses, we consider only the number of citations received after two years of publication. The findings indicate the following trends: (a) for sole-authored papers, male authors receive a slightly higher average number of citations (3.49) compared to female authors (3.36). The p-value of 0.02 suggests that this difference is statistically significant, indicating that sole-authored papers by male authors tend to receive more citations than those by female authors; (b) In two-author teams, the average number of citations for male-only (4.62) and female-only (4.20) teams are relatively close, with a p-value of 0.66, indicating no statistically significant difference; (c) For three-author teams, male-only teams receive significantly more citations on average (5.32) compared to female-only teams (4.12), with a p-value of 0.0004, indicating a statistically significant disparity. However, when comparing teams with more male authors to those with more female authors within this group, the average number of citations (4.85 vs. 4.56) shows no significant difference, as indicated by a p-value of 0.11; (d) In four-author teams, male-only teams receive slightly more citations on average (4.91) compared to female-only teams (4.37), but the difference is not statistically significant (p-value = 0.64). Similarly, for teams with more male or more female authors, the average number of citations (5.02 vs. 4.74) shows no significant difference, with a p-value of 0.17; (e) For five-author teams, male-only teams again receive more citations on average (5.36) compared to female-only teams (4.66), but the difference is not statistically significant (p-value = 0.36). The comparison between teams with more male authors and those with more female authors (5.44 vs. 4.91) also shows no significant difference, as indicated by a p-value of 0.36.

To summarize, we find that sole-authored papers by male authors tend to receive significantly more citations on average compared to those by female authors. This gender disparity in citations is also evident in three-author teams, where male-only teams receive significantly more citations than female-only teams. However, within teams larger than 5 authors, particularly those with a mix of male and female authors, the differences in citation counts are not statistically significant. This may suggest that gendered citation disparities are salient in sole or small-group authorship, while larger teams receive more balanced outcomes regardless of gender composition.

\section{Most Influential Journals in Communication}

In addition to analyzing gender composition in authorship, we also focus on the most prestigious journals that shape and represent the discipline. These journals often set the standards for research quality and influence the direction of scholarly discussion. Therefore, understanding gender representation within these top-tier publications can reveal critical patterns and disparities that might not be apparent in a broader analysis.

To address this, we conducted a detailed investigation into the gender composition of authorship within the top 10 flagship communication journals. This analysis covers a total of 7,017 publications from 2004 to 2023. By focusing on these leading journals, we aim to uncover whether women are equally represented in the spaces that are most visible and influential within the field. The journals selected for this analysis are recognized for their impact and contribution to the advancement of communication research. They include {Communication Methods and Measures}, {Political Communication}, {Communication Research}, {Digital Journalism}, {Journal of Advertising}, {Journal of Communication}, {Journal of Interactive Advertising}, {Annals of the International Communication Association}, {International Journal of Press-Politics}, {Big Data \& Society}, {Public Opinion Quarterly}, {New Media \& Society}, and {International Journal of Advertising} (see Table \ref{fig_top}). Examining gender representation in these key publications provides a focused lens through which to assess the progress and remaining challenges regarding gender equity in the field of communication research.

The analysis of papers in these journals reveals important insights into gender disparities in academic publishing and citation impact, as shown in Table \ref{fig_top}. These journals are listed in descending order of the percentage of female authorship, providing a clear view of how gender representation correlates with citation metrics. The metrics analyzed include the number of publications, impact factor (IF), percentage of female authors, and the differences in citation percentages between male and female authors under various authorship conditions.  Across these journals, the percentage of female authors varies from 33.1\% to 47.0\%, indicating that while some journals approach gender parity, women are still underrepresented (i.e., the proportions of female authors were lower than 50\%). Particularly in \textit{Public Opinion Quarterly}, female authors constitute only 33.1\% of the total.

The Impact Factor (IF) of these journals ranges from 3.4 to 11.4, with \textit{Communication Methods and Measures} standing out with the highest IF of 11.4. Notably, this journal also exhibits the most significant disparity in citation impact between male and female sole authors, with female-authored papers receiving 107.59\% more citations than those authored by males. In the case of sole authorship, female researchers received more citations than their male colleagues in 7 out of 10 journals. 

Despite more citations received by sole female authors, female authors still receive fewer citations compared to males when it comes to multiple-author papers. First, female first-authored papers generally receive fewer citations. For instance, in Political Communication, female first-authored papers receive 25.32\% fewer citations. Similarly, \textit{Digital Journalism} and\textit{ Communication Methods and Measures} also exhibit substantial negative differences, with female first-authored papers receiving 19.86\% and 76.97\% fewer citations, respectively. Teams composed entirely of female authors also tend to receive fewer citations compared to all-male teams. For example, in \textit{Communication Methods and Measures}, all-female teams receive 82.07\% fewer citations than all-male teams. Disparities are also evident in Public Opinion Quarterly and Political Communication, where all-female teams receive 36.81\% and 27.8\% fewer citations, respectively. Teams with a higher proportion of female authors also receive fewer citations; in \textit{Political Communication} and Communication Methods and Measures, such teams report negative differences of 36.0\% and 78.4\%, respectively. This stark contrast across team sizes underscores the uneven recognition of male and female scholars within high-impact venues. 

A notable observation from the data is the difference in citation patterns between sole-authored papers and multiple-authored papers. In many cases, female authors publishing alone tend to receive more citations than their male counterparts. However, this trend reverses when examining multiple-authored papers. Teams composed entirely of female authors or teams with more female authors generally receive fewer citations than their male counterparts. This discrepancy suggests that while individual female authors may achieve notable recognition, the collaborative dynamics within academic publishing might disadvantage female-dominated teams. The reasons behind this could be multifaceted, including biases in citation practices, the visibility and networking opportunities available to male versus female scholars, and possibly the perception of the contributions of team members based on gender composition.
  
On the other hand, there are a few exceptions where female authors or female-dominated teams receive more citations. In the \textit{Journal of Advertising}, teams with more female authors receive 26.4\% more citations compared to teams with more male authors. \textit{Communication Research} also shows a favorable difference for female first-authored papers (13.21\%) and for sole-authored papers (8.78\%).

In addition, we also examined the probability of forming teams with varying gender compositions and the corresponding citation impact in these prestigious journals. The analysis reveals that male-only and female-only teams are more prevalent than mixed-gender teams, particularly when team size increases. For instance, in four-author teams, the probability of it being a male-only team is 32.16\%, and for a female-only team, it is 99.61\%. This pattern of homophily — where individuals tend to collaborate with others of the same gender — is dominant, especially in larger teams.

However, despite the higher probability of male-only and female-only teams, the average number of citations received by these teams does not differ significantly.  Results show no statistically significant differences in citation rates between male-only and female-only teams regardless team sizes. This suggests that while gender may influence team formation, other factors may also parse the scholarly impact of the research, as measured by citations.

Overall, these findings highlight the persistent gender disparities in the communication field, particularly in the composition of research teams and the citation impact of their work. The trend of gender homophily in team formation interacting with uneven recognition of male and female scholars, pinpoint existing biases in the academic community. Addressing these imbalances requires further investigation into their underlying causes and initiatives aimed at supporting female researchers in gaining equitable recognition. This calls for concerted efforts from journals, academic institutions, and the broader scholarly community to ensure that female scholars in communication receive the same opportunities and acknowledgment as their male counterparts.

\section{Gender Representation and Salary Distribution in Communication Departments Among the Top U.S. Public Universities}

Understanding faculty compensation within leading communication departments is critical for addressing systemic inequities and fostering a fair academic environment. We initiated a Public Records Act request to obtain detailed salary information across various faculty ranks — Assistant Professors, Associate Professors, and Full Professors —  at prominent public universities in the United States. This data collection effort serves a dual purpose: a thorough analysis of salary disparities and promotion of equitable pay structures within academia. Given the influential role of these institutions, their compensation practices have weighty implications for the field of communication studies. By examining these practices, we aim to provide valuable insights for policymakers and university administrators in evaluating and calibrating salary structures with broader institutional goals and industry standards \cite{cahalan2022indicators, copeland2020diversity}. Transparency in faculty compensation not only enhances accountability but also empowers stakeholders to make informed decisions, ultimately contributing to a more equitable and inclusive academic landscape.

The selected institutions were chosen based on their prominence and influence within the field of communication studies, ensuring that the sample represents a diverse range of leading academic environments. As such, we seek to obtain a representative overview of compensation practices across these institutions. Table \ref{table_number} offers a detailed view of the gender distribution among faculty members at American Universities, categorized by academic rank—Assistant Professor, Associate Professor, and Professor. The data reveal significant variability in female representation across universities and ranks. For Assistant Professors, the percentage of women ranges dramatically, from 0\% at the University of California - Davis to 100\% at the University of Arizona. Overall, 61.42\% of Assistant Professors are female. The variation is even greater at the Associate Professor level, where female representation spans from 25\% at Michigan State University to 70.37\% at the University of Texas at Austin. On average, 51.64\% of Associate Professors are female. At the rank of (full) Professor, the percentage of women varies from 16.67\% at the University of Arizona to 87.50\% at the University of Washington, with an overall representation of 47.26\%.

Findings indicate that in the field of communication while female faculty members are well-represented at the Assistant Professor level, their proportion tends to decrease at higher academic ranks. This trend suggests that women are faced with challenges when advancing to higher ranks despite visible progress in female representation in early career stages. The data also indicate institutional differences in gender representation. For example, the University of Texas at Austin exhibits a relatively balanced female representation, whereas institutions like Michigan State University and the University of Arizona display more uneven distributions. Overall, the data show advancement but also challenges in achieving gender parity within academic faculty. The lower percentage of women in higher ranks implies hurdles to career progression for female faculty, which warrants further investigation and efforts to support and retain women in academia as they advance through their careers.

Furthermore, we examined average salaries of male and female faculty members at the selected American universities, categorized by academic rank: Assistant Professors, Associate Professors, and Full Professors, as presented in Table \ref{table_salary}. For each university, we present the average salary for both genders at each rank and calculate the percentage difference in salaries between male and female faculty. For Assistant Professors, we observe considerable variability in salary differences across universities. For instance, at Purdue University, female Assistant Professors earn 24.87\% less than their male counterparts, which is the most notable disparity in the data. On the other hand, at the University of Wisconsin-Madison, female Assistant Professors earn 8.02\% more than their male peers. On average, female Assistant Professors earn about 5.61\% less than their male colleagues across the institutions surveyed. This trend suggests a tendency for female Assistant Professors being paid less than their male counterparts, though the extent of this disparity varies by institution.

When it comes to Associate Professors, the percentage differences in salaries are generally smaller compared to Assistant Professor, but varied institutionally. Female Associate Professors earn slightly more than their male colleagues, with an overall difference of +2.61\%.  At Purdue University, female Associate Professors earn 14.75\% less than their male counterparts, while at the University of California - Davis, female Associate Professors earn 27.27\% more.  To conclude, while there are instances of lower salaries for female Associate Professors at certain universities, the average difference across institutions suggests a slight advantage for females in this rank.

For (Full) Professors, the average salary difference is minimal, with only a -0.03\% disparity, indicating that salaries between male and female professors are nearly equivalent. At the University of Michigan, female Professors earn 8.28\% more than their male counterparts. However, at Purdue University, male Professors earn 6.03\% more than female Professors. 

Altogether, the results illustrate a nuanced picture of gender-based salary disparities in the field of communication. Assistant Professors, in general, face a salary disadvantage compared to their male peers, with an overall 5.61\% lower salary. For Associate Professors, female faculty members tend to earn slightly more on average, with a 2.61\% advantage. Salaries for (Full) Professors are almost equal, reflecting a negligible difference of -0.03\%. These findings suggest that gender-based salary disparities persist but vary by academic rank and institution. Addressing these disparities is critical to achieving gender-equitable compensation in academia.

\section{Conclusions}

The present study examines gender dynamics within communication research and demonstrates a complex landscape informed by citation metrics, authorship patterns, team composition, and salary disparities. Our findings underscore critical trends in gender influences academic recognition, collaboration, and compensation.

Citation metrics remain the most essential component of evaluating research impact, often affecting funding decisions and career advancement. Our findings highlight a nuanced reality: sole-authored papers by female researchers receive fewer citations on average compared to their male counterparts, though this disparity narrows in larger, more diverse teams. This may suggest that although gender-related citation biases are evident, they could be mitigated by collaborative efforts and diverse research teams. In addition, our findings on team composition reveal a tendency towards gender homophily, with single-gender teams being more prevalent than mixed-gender teams. This pattern of collaboration, while prevailing, does not consistently affect the citation impact of research. With that said, it indicates underlying biases in team composition and highlights the need for more inclusive and diverse research environments.

Our examination of authorship in the top communication journals provides further evidence for the gender disparities in academic publishing. Despite some progress, female authors remain underrepresented, particularly in high-impact journals. The significant citation disparities between male and female sole authors in certain journals reflect ongoing challenges in achieving equitable recognition for female scholars.

Salary analysis within top U.S. public universities reveals that while female faculty members are well-represented at entry levels, the proportion of females decreases at higher academic ranks. Along with the observed salary differences, this disparity underscores the need for addressing career progression barriers and ensuring fair compensation, calling for systematic and institutional efforts. Although disparities in salaries are minimal at higher ranks, further vigilance is required to foster equity across all levels. To summarize, thoroughly addressing gender disparities in communication research requires a multifaceted approach. Promoting diversity in collaboration, ensuring equitable citation practices, and addressing salary disparities are crucial steps toward achieving gender equity and eliminating gender disparities in academia. Continued efforts from journals, academic institutions, and the broader scholarly community are key to supporting and advancing female researchers, ensuring that all scholars have equal opportunities to contribute to and be recognized in the field of communication.



\end{document}